# Thermal conductivity of PbTe-CoSb$_3$ bulk polycrystalline composite: role of microstructure and interface thermal resistance

Artur Kosonowski[a], Ashutosh Kumar[b], Taras Parashchuk[b], Raul Cardoso-Gil[c], Krzysztof T. Wojciechowski[a*]

[a]AGH University of Science and Technology Faculty of Materials Science and Ceramics Mickiewicza ave. 30, 30-059 Krakow, Poland

[b]The Lukasiewicz Research Network - Cracow Institute of Technology, Zakopianska 73, 30-418 Krakow, Poland

[c]Max-Planck-Institut für Chemische Physik fester Stoffe, Nöthnitzer Str. 40, 01187 Dresden, Germany

*corresponding author: wojciech@agh.edu.pl

ABSTRACT

Systematic experimental and theoretical research on the role of microstructure and interface thermal resistance on the thermal conductivity of the PbTe-CoSb3 bulk polycrystalline composite is presented. In particular, the correlation between the particle size of the dispersed phase and interface thermal resistance ($R_{int}$) on the phonon thermal conductivity ($\kappa_{ph}$) is discussed. With this aim, a series of PbTe-CoSb3 polycrystalline composite materials with the different particle sizes of CoSb$_3$ was prepared. The structural (XRD) and microstructural analysis (SEM/EDXS) confirmed assumed chemical and phase compositions. The acoustic impedance difference ($\Delta Z$) was determined from measured sound velocities in PbTe and CoSb$_3$ phases. It is shown that the $\kappa_{ph}$ of the composite may be reduced when the particle size of the dispersed phase (CoSb$_3$) is smaller than the critical value of ~230nm. This relationship was concluded to be crucial for controlling the heat transport phenomena in composite thermoelectric materials. The selection of the components with different elastic properties (acoustic impedance) and particle size smaller than the Kapitza radius leads to a new direction in the engineering of composite TE materials with designed thermal properties.

1. INTRODUCTION

According to recent reports, the majority of energy (~65%) generated across all areas of industry is wasted or rejected [1]. Thermoelectric (TE) materials are suitable for the conversion of waste heat into electrical energy and find their applications in the automotive industry [2,3], for power supply during space exploration [4] or in the internet of things [2,5]. The efficiency of energy conversion in TE materials depends on a dimensionless quantity known as the TE figure of merit ($ZT$)

$$ZT = S^2 \sigma \kappa^{-1} T \qquad (1)$$

where $\sigma$ is the electrical conductivity, $S$ is the Seebeck coefficient, $\kappa$ is the thermal conductivity, and $T$ is the absolute temperature. The total thermal conductivity ($\kappa$) consists of electronic contribution ($\kappa_e$) and phonon thermal conductivity ($\kappa_{ph}$) i.e., $\kappa = \kappa_e + \kappa_{ph}$.

Lowering of $\kappa_{ph}$ is one of the primary requirements to achieve a high $ZT$ in the TE materials and has been demonstrated in literature through various strategies that enhance the phonon scattering, including nanostructuring [6,7], lattice defects [8,9], artificial superlattices [10], mass disorder [11], the introduction of nano-micro-porosity [12] and by preparation of composite materials [13–16]. However, phonon scattering via lattice defects, impurities, and lattice anharmonicity in a single-phase TE material may also enhance the scattering of electrons that results in a lower charge carrier mobility and hence reduced $\sigma$ [17].

In particular, the preparation of hetero-phase systems made of two conductive components or semiconductors can lead to reducing $\kappa_{ph}$ without deteriorating the electrical conductivity $\sigma$. As an example, an improvement of TE properties was observed for PbTe- PbSe composite due to the presence of nano-precipitates obtained during spinodal decomposition [18]. The authors of this work achieved almost 40% reduction in $\kappa_{ph}$ (~0.4 Wm$^{-1}$K$^{-1}$), and a high power factor $S^2\sigma$ due to optimization of electronic properties and in result high $ZT$ (~1.85). A significant reduction in $\kappa_{ph}$ (~30%) was shown in CoSb$_3$-ZrO$_2$ nanocomposite with the increase in amount of nano-sized zirconia [19]. Katsuyama *et al.* investigated the TE properties in CoSb$_3$-oxide dispersed in MoO$_2$, WO$_2$, and Al$_2$O$_3$, and they observed a lowering of $\kappa_{ph}$ when the grain size of both phases decreased [20].

Most of the authors usually attribute lowering $\kappa_{ph}$ in composite TE materials to quantum size effects (i.e., the influence of nano-inclusions) [15,18–20]. These studies on hetero-phase materials are focused mostly on general transport properties and rarely refer to the role of interface thermal resistance ($R_{int}$), and acoustic impedance ($Z$) on total thermal conductivity of investigated systems [24–26]. However, these factors are often used in the explanation of heat transport mechanisms in ceramic and polymer composites (e.g. diamond/ZnS [27], SiC/Al [28,29], diamond/cordierite [30,29], alumina/epoxy, and glass/epoxy [31]). In our opinion, this approach did not receive significant attention in the development of efficient thermoelectric materials.

We present the results of systematic studies on thermal transport in the model system, composed of two semiconducting phases, using effective media theory (EMT) that considers the elastic properties of the phases in terms of $R_{int}$ and $Z$. For this aim the model system should fulfil the following requirements: (a) components should show excellent thermoelectric performance (high ZT), (b) both materials need to have the same type of majority carriers (*n*- or *p*-type semiconductor), (c) materials can not react with each other over the investigated temperature range, (d) the number of elements in the composite should be kept at a minimum (e.g. doping by deviation from stoichiometry or elements used for doping should be already present in the composite). These assumptions aim to avoid the interaction of phases during thermal treatment, which can result in the formation of impurity phases or diffusion of elements between phases [31,32]. As a result of our analysis of the chemical and physical properties of many compounds, we have selected two well-known and attractive TE materials: PbTe and $CoSb_3$. Both the compounds were also reported as ingredients in composite systems[18–20,32]. However, only a few studies on $CoSb_3$ (as a matrix) and PbTe (as nano-inclusions up to 8 wt%)[33,34] composite are shown in the literature. The transport properties of the composite system were attributed to the nanostructuring effects and the authors did not consider the influence of microstructure and interface thermal resistance. Hence it seems to be crucial to study the thermal transport in the (1-x)PbTe/(x)$CoSb_3$ composite across a wide composition range (0 ≤ x ≤1.00) considering the role of particle size of the dispersed phase and interface thermal resistance between the phases.

PbTe crystallizes in an *F*m-3m space group (NaCl-type structure) [35], where Pb and Te atoms are coordinated octahedrally. It is a narrow bandgap semiconductor ($E_g$ = 0.19eV at 0K [36]) used for thermoelectric applications at temperatures 500-900K [37]. It has low carrier effective mass ($m≈0.01m_e$), resulting in their high mobility [35]. Its unique electronic properties lead to high *ZT* values over 2 at temperatures above 700K [38]. By introducing acceptor or donor impurities PbTe can be easily modified for *p*- or *n*-type of conductivity [39,40]. In our case, we propose to use Sb as a donor dopant to achieve n-type conductivity [40]. The highest ZT ~1.1 is obtained for $Pb_{0.99}Sb_{0.01}Te$ composition [41].

On the other hand, $CoSb_3$ also possesses a cubic structure (*I*m-3) with two large structural voids per unit cell [42], and like PbTe it is a narrow bandgap semiconductor ($E_g$ = 0.22eV at 0K [43]). Pristine $CoSb_3$ shows an attractive value of the Seebeck coefficient (>200 µV/K) due to the large effective mass of carriers ($m≈0.18m_e$) and relatively high electrical conductivity (~900S/cm at 25°C [44]). However, for obtaining a high ZT parameter, proper doping or filling of structural voids is often used to reduce the lattice thermal conductivity. The highest reported values of ZT are ~1.9 above 800K [45] for $Sr_{0.09}Ba_{0.11}Yb_{0.05}Co_4Sb_{12}$. For this research, we used Te as a dopant, and to obtain *n*-type $CoSb_3$ we applied moderate sintering pressure to avoid filling the structural voids at 2a Wyckoff position by Te atoms[46]. We chose the composition $CoSb_{2.94}Te_{0.06}$, which represents the solubility limit of this element in the $CoSb_3$ structure [47,48] which should minimalize diffusion of Te from the PbTe phase. For this amount of Te, the highest reported ZT is ~1.0 [49].

We expect that the composites of both selected components will be chemically stable. We plan to fabricate of series (1-*x*)$Pb_{0.99}Sb_{0.01}Te$/(*x*)$CoSb_{2.94}Te_{0.06}$ (hereinafter referred to as (1-x)PbTe/(x)$CoSb_3$) composite samples with different grain size and amount *x* of $CoSb_{2.94}Te_{0.06}$ and characterize their thermal properties – as presented schematically on Figure 1. Using EMT theory, we explain the influence of the interface thermal resistance $R_{int}$ and microstructural parameters (Kapitza radius $a_K$) on the phonon thermal conductivity of the composites. Finally, we demonstrate the criteria for the development of effective composite thermoelectric materials with as low as possible thermal conductivity.

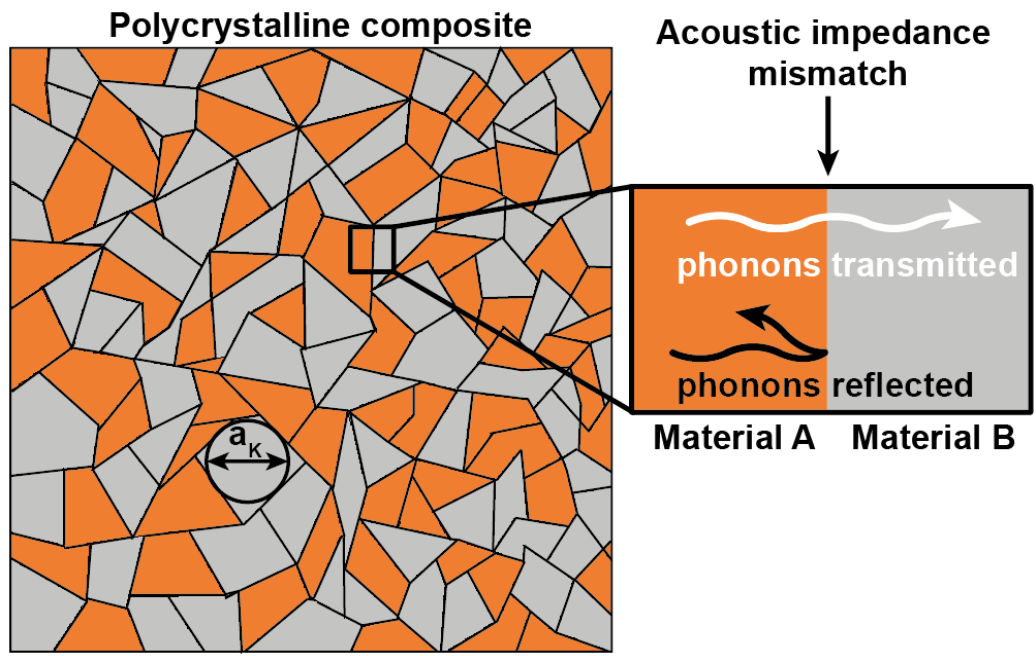

**Figure 1.** Schematic of polycrystalline composite material and physical parameters that influence phonon thermal conductivity of the material ($a_K$ – Kapitza radius).

2. EXPERIMENTAL SECTION

Synthesis of the PbTe and CoSb$_3$ was carried out by direct melting of elements with purity 99.999% (Alfa Aesar) in evacuated quartz ampoules. A furnace with a rocking mechanism was used to provide homogenization of the melted elements. The materials were heated up to 1273 K in 5 hours and kept at this temperature for one hour and then cooled down through radiative heat loss to 1123 K for 1 hour. Further, PBTE ampoule was taken out from the furnace at 1123 K and quenched in the air. The CoSb$_3$ ampoule was kept at 1123 K for seven days, followed by natural cooling to the room temperature. Ingots of both materials were separately milled (PM100, RETSCH) or hand ground to obtain a batch of PbTe with ~10 μm particle size distribution and three batches of CoSb$_3$, each with different particle size distribution (~300 μm, 30 μm, and 1 μm). Milling parameters for each case are presented in Supplementary Information (Table S1). Obtained powders were subjected to the analysis of the particle size distribution by the laser light diffraction method (Mastersizer 2000S, Malvern Instruments). Powders were mechanically mixed and then sintered to obtain the (1-$x$)PbTe/($x$)CoSb$_3$ composite with 0≤$x$≤1 ($x$ being the volume ratio) with 30 μm particle size of CoSb$_3$ phase. We have also prepared two composite materials (0.5)PbTe/(0.5)CoSb$_3$ with a mode particle size of 300 μm and 1 μm of CoSb$_3$ (Fig.2). The volume ratio $x$ was used for the calculation of the amount of components weighted according to the formula: $x \cdot v_c \cdot \rho = m$, where $v_c$ is the volume of the composite sample, $\rho$ is the crystallographic density of the CoSb$_3$ phase, and $m$ is the mass of the CoSb$_3$ phase required for the preparation of a given composite sample. The composite mixtures were sintered using Pulsed Electric Current Sintering (PECS) technique in Ar (5N) atmosphere under the following conditions: heating rate 100 K/min, sintering temperature 973 K maintained for 15 min, uniaxial pressure 50 MPa applied at 973 K, and released at the beginning of the cooling segment (30 K/min). Obtained cylindrical samples with 10 mm diameter and 15 mm length were cut to proper dimensions for further measurements. The special layered sample (Fig. 5a) for investigation of the interface thermal resistance ($R_{int}$) between PbTe and $CoSb_3$ phase was prepared by putting the powder layer of each material in a die with a 6 mm diameter and applying the same sintering procedure as described above. Obtained layers had a thickness of ~1 mm each. All sintered pellets were subsequently annealed in Ar-filled quartz ampoules at 823 K for 20 hours. For the investigation of the microstructure, the surface of the samples was polished using an automatic grinding/polishing machine. Then, the surface was chemically etched for ~5 seconds with Murakami reagent (solution of potassium ferricyanide in sodium hydroxide). SEM analysis of composite samples was performed using the NOVA NANO SEM 200 (FEI EUROPE COMPANY) microscope equipped with an EDXS analyzer. Microstructure analysis of the layered composite sample was performed using optical microscopy (Axioplan 2, Carl Zeiss, polarized light) and SEM/EDXS analysis (JEOL 7800F with an attached EDX/EBSD system: Quantax 400). X-ray diffraction of samples after synthesis and after sintering was obtained by the D8 ADVANCE (BRUKER) diffractometer using Bragg-Brentano geometry and Ni-filtered Cu-Kα radiation (λ=1.5406 Å) in the 2θ range 15-85°. The Rietveld refinement for all materials was performed using GSAS II software [50].

Thermal diffusivity for all samples was measured using the laser flash analysis (LFA 457, NETZSCH) in Ar (5N) atmosphere (30 ml/min). Specific heat was determined simultaneously with the thermal diffusivity using Pyroceram 9606 (NETZSCH) as reference material. The sample density was calculated from the measurements of the sample mass and its geometrical volume. The porosity obtained from the calculation of relative density was taken into account during the analysis of thermal conductivity for all materials using porosity correction formula [51]. Sound velocity was measured at room temperature using ultrasonic flaw detector EPOCH 3 (PANAMETRICS). All measurements were performed in the direction parallel to the pressing force. The estimated uncertainty of the thermal diffusivity results is 5%, thermal conductivity 7%, and interface thermal resistance about 10%.

3. RESULTS AND DISCUSSION

First, all prepared materials were subjected to the analysis of structure, phase composition, microstructure, and chemical composition using XRD and SEM/EDXS methods. Additionally, a special layered sample was fabricated for measurement of the interface thermal resistance using the LFA method, and the boundary between layers was investigated by optical and electron microscopy.

## 3.1 STRUCTURAL AND MICROSTRUCTURAL ANALYSIS

The particle size distribution of powders before sintering is presented in Fig. 2. The application of the same milling conditions for PbTe and $CoSb_3$ (black bars Fig. 2(a) and (b)) leads to comparable particle sizes of materials in the range from about 0.2 μm to about 200 μm. Small differences in the location of the distribution modal value may be due to the difference in the hardness of $CoSb_3$ [52] and PbTe [53]. Also, different milling conditions applied to $CoSb_3$ produced three different ranges of particle size distributions, as shown in Fig.2(b). A series of $(1-x)PbTe/(x)CoSb_3$ with $0 \leq x \leq 1$ composite material was prepared using similar average particle size of PbTe (~10 μm) and $CoSb_3$ (~30 μm). Further, two composite materials $(0.5)PbTe(0.5)CoSb_3$ were also prepared with the same particle size of PbTe (~10 μm) but with different particle sizes of $CoSb_3$ (~300 and 1 μm respectively).

Structural characterization of the $(1-x)PbTe/(x)CoSb_3$ composite with $0 \leq x \leq 1$ was performed on sintered samples using the powder X-ray diffraction (XRD) method. The analysis of all prepared materials confirmed the presence of characteristic reflections of PbTe and $CoSb_3$ only; no impurity phases were observed. The intensity corresponding to $CoSb_3$ reflections was found to increase proportionally with the increase in the $CoSb_3$ phase fraction in the composite, as shown in Fig. 3(a).

Further, a quantitative analysis of the XRD pattern was done using Rietveld refinement. The refined patterns for the samples with x=0, 0.5, 1 are shown in Fig. 3(b) and the refinement parameters for all the samples are listed in Table I. The weight fractions (converted to volume fractions) determined from the refinement agree with the nominal compositions within experimental error. Refined patterns for $(0.5)PbTe/(0.5)CoSb_3$ with $CoSb_3$ particle size of 300 μm and 1 μm are in the Supplementary Information (Fig. S1)

The experimental, calculated and relative densities, are presented in Table I. The experimental density for all the samples with $CoSb_3$ particle size ~30 μm is changing monotonically according to the rule of mixtures. The relative density is higher than 98%.

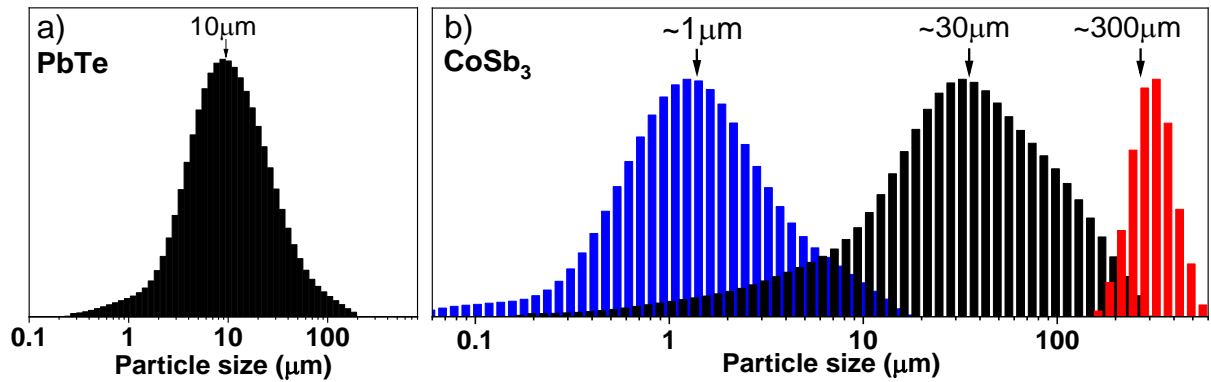

**Figure 2.** The particle size distributions of (a) PbTe, and (b) $CoSb_3$ powders (modal values are shown).

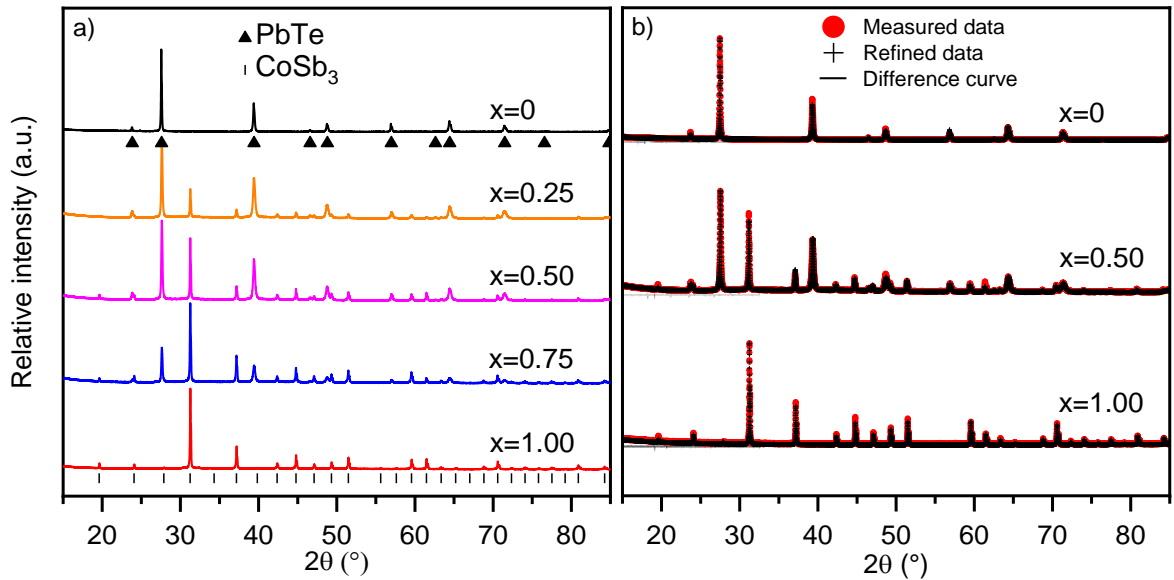

**Figure 3.** (a) Powder x-ray diffraction pattern (Cu-K$\alpha$ radiation) for polycrystalline (1-x)PbTe/(x)CoSb$_3$ composite after PECS sintering. The Bragg's position for both PbTe and CoSb$_3$ phases are marked, (b) Rietveld refinement patterns for (1-x)PbTe/(x)CoSb$_3$ composite for x=0, 0.50 and 1.00.

Microstructure and chemical composition of sintered pellets have been carried out on polished-etched cross-sections and the fracture surface, as shown in Fig. 4(a,b). The two separate phases without any sign of the third phase at the grain boundary are observed in the SEM images, which agrees with the conclusion from the XRD analysis. Figure 4(a,b) shows that particles of two phases are uniformly distributed across the investigated area, and the size of particles lies in the range from 5 $\mu$m to 50 $\mu$m. It agrees with the particle size distribution after ball milling for both materials (Fig. 2). Additional phase fraction analysis was performed on the SEM image shown in Fig. 4(a) using ImageJ software [54]. This analysis reveals that the volume fraction for PbTe and CoSb$_3$ are 0.53 and 0.47, respectively, which is in good agreement with the values obtained from Rietveld refinement (Table I). Figure 4 (c,d) represents the surface of the fractured samples and shows a specific area, where a significant amount of smaller PbTe particles is observed. In between those grains, micro-porosity is present and is consistent with the relative density obtained for these materials (Table I).

**Table I:** Composition of the (1-$x$)PbTe/($x$)CoSb$_3$ composites from Rietveld refinement

| Composition | Nominal $x$ | Experimental $x$ (±0.05) | $R_{wp}$ [%] | GOF | Density Experimental [g/cm$^3$] | Density Relative [%] |
|---|---|---|---|---|---|---|
| PbTe 10μm + CoSb$_3$ **300μm** | 0.50 | 0.55 | 10.167 | 3.21 | 7.69 | 97.4 |
| PbTe 10μm + CoSb$_3$ **30μm** | 0.00 | 0.00 | 8.025 | 2.11 | 7.98 | 97.3 |
| | 0.25 | 0.30 | 8.363 | 2.73 | 7.82 | 97.1 |
| | 0.50 | 0.45 | 7.539 | 2.53 | 7.73 | 97.7 |
| | 0.75 | 0.73 | 5.506 | 1.94 | 7.65 | 98.5 |
| | 1.00 | 1.00 | 8.182 | 1.26 | 7.26 | 95.2 |
| PbTe 10μm + CoSb$_3$ **1μm** | 0.50 | 0.52 | 11.869 | 3.87 | 7.36 | 93.2 |

For a detailed investigation of the interface between PbTe and CoSb$_3$ phases in the composite, an additional layered sample was prepared, as mentioned before (Fig. 5(a)). Linear EDXS analysis has been performed across the boundary between layers of PbTe and CoSb$_3$, which is shown in Fig. 5(b) by a red dashed line. It is observed that under given magnification, the amounts of elements are almost constant within each layer, and a step-like change in the chemical composition can be observed at the interface. The width of the observed transitional region is ~1 μm, which is very small as compared to the particle sizes of individual phases. Hence, we conclude that there is no sign of significant chemical diffusion at the interface between the PbTe and CoSb$_3$ phases.

Further, the optical microscopy image of the layered sample in polarized light is shown in Fig. 6(a). A clear separation between two phases at the interface is seen, which is further supported by the SEM images in Fig. 6(b). Additionally, the EDXS mapping for the elements present in the layered sample is performed at the same location, and corresponding images are shown in Fig. 6(c-f). It is seen that the transition is sharp, and the elements are distinguishable at the interface. It also confirms that the composite prepared in the present study is chemically stable and does not show noticeable chemical diffusion at the interface.

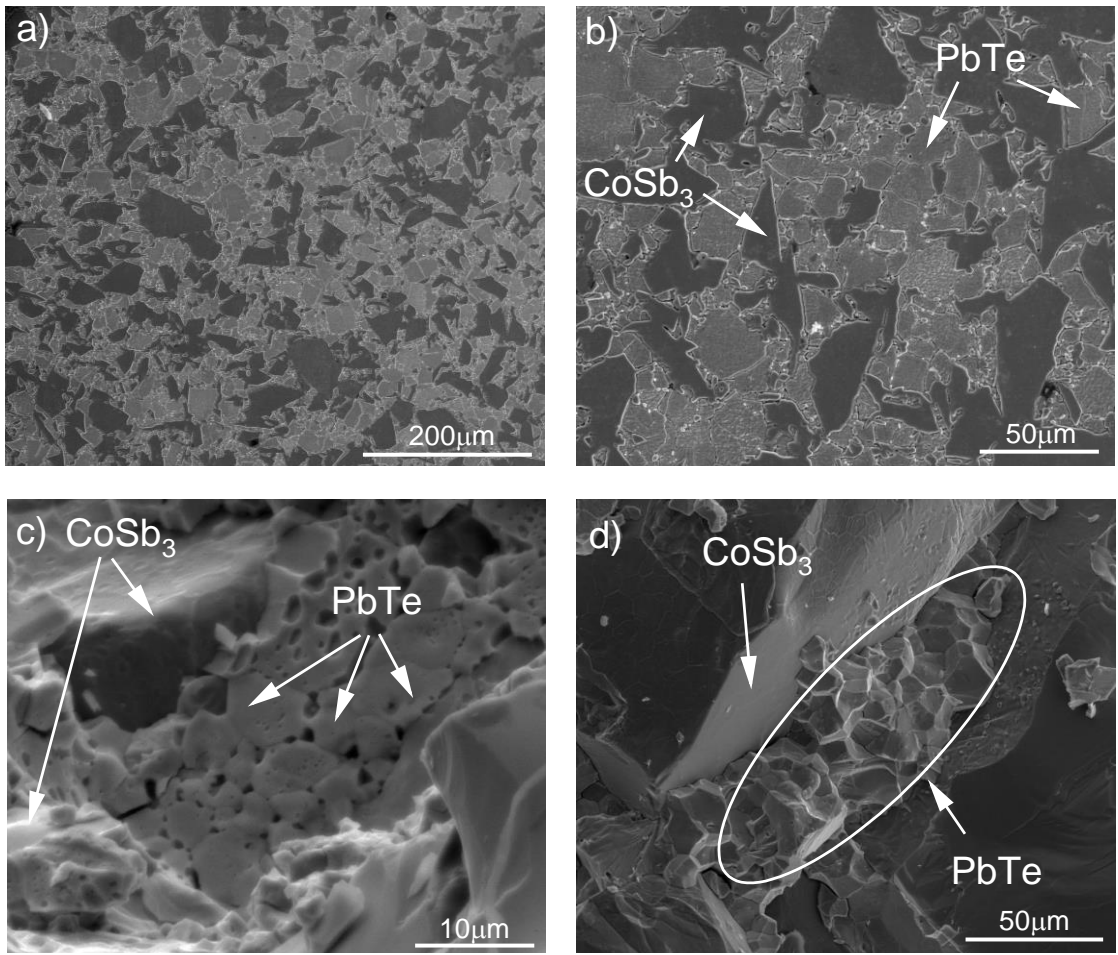

**Figure 4.** Scanning electron microscope (SEM) images of (a,b) polished/etched surface, (c) fractured surface for (0.50)PbTe/(0.50)CoSb$_3$ composite having 30μm particle size of CoSb$_3$, (d) fractured surface of (0.50)PbTe/(0.50)CoSb$_3$ sample having 300μm particle size of CoSb$_3$.

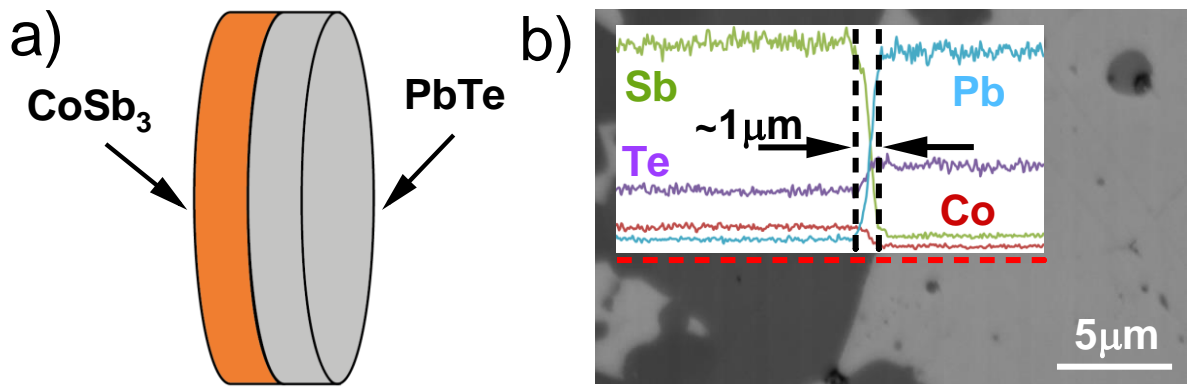

**Figure 5.** (a) Scheme of the PbTe-CoSb$_3$ layered sample, (b) linear EDXS analysis of the interface between PbTe and CoSb$_3$ phases (red dashed line indicates the measured line).

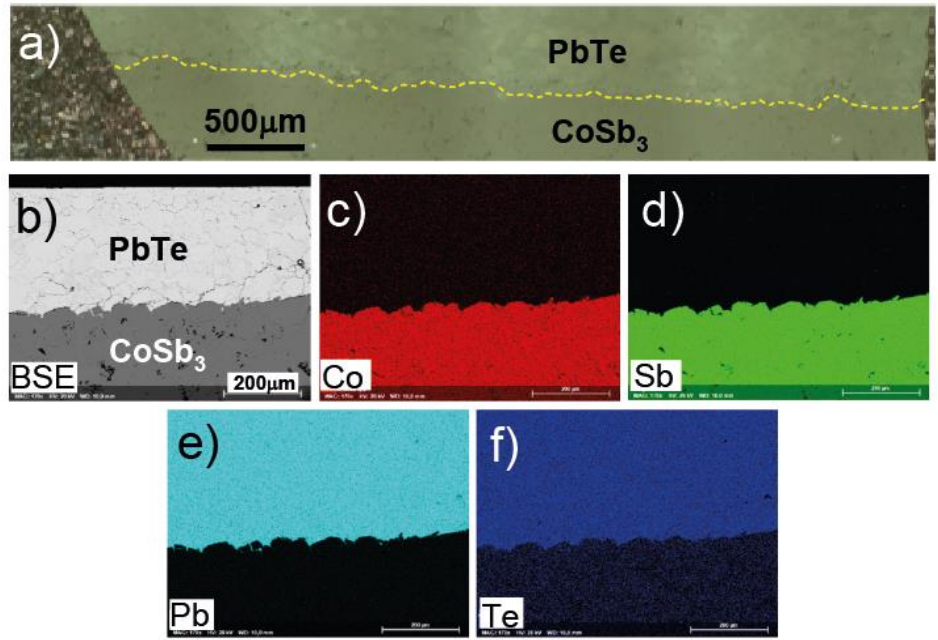

**Figure 6**. (a) picture from the optical microscope with polarized light of the interface, (b) SEM image and (c-f) EDXS mapping of the area near the interface between PbTe and $CoSb_3$ layers.

## 4. THERMAL PROPERTIES

After confirmation of the binary phase composition of prepared composites, their thermal properties are investigated. Also, the results of these measurements were compared with theoretical values obtained using the effective medium theory. Further, the correlation between the interface thermal resistance and the particle size of the dispersed phase on the thermal conductivity of the composite is discussed.

### 4.1 THERMAL CONDUCTIVITY

The total thermal conductivity ($\kappa$) of (1-x)PbTe/(x)$CoSb_3$ for 0≤x≤1 composite samples is presented in Fig.7(a) over a temperature range from 50°C to 400°C. The $\kappa$ for PbTe at 50 °C is ~2 $Wm^{-1}K^{-1}$ and ~6 $Wm^{-1}K^{-1}$ for $CoSb_3$ at the same temperature. For all the prepared materials, the $\kappa$ decreases with the increase in temperature. Further, for samples with x=0.00, 0.25, and 0.50 total thermal conductivity has similar values at lower temperatures. However, at higher temperatures, a monotonous change in $\kappa$ with the increase in the $CoSb_3$ phase fraction is observed for all samples. Additionally, $\kappa$ of composite with different particle sizes of the $CoSb_3$ phase is presented in Fig. 7(c). The $\kappa$ for the composite with 300 μm and 30 μm particle size of $CoSb_3$ shows similar values within the experimental error. However, the $\kappa$ values for the composite with 1 μm particle size of $CoSb_3$ show minimal reduction compared to the other two samples, as shown in Fig.7(c).

For estimation of phonon thermal conductivity $\kappa_{ph}$ in the PbTe and $CoSb_3$ systems, we applied Wiedemann-Franz law to calculate the electronic thermal conductivity ($\kappa_e = L\sigma T$) and subtracted

$\kappa_e$ from total thermal conductivity. The temperature dependence of the Lorenz numbers $L_{PbTe}$ and $L_{CoSb3}$ was calculated using the Kane model, described in more detail in the supplement to the article. The Lorenz numbers for composites $L_c$ were estimated using the rule of mixtures based on calculated $L_{PbTe}$ and $L_{CoSb3}$ values (Fig. S2). Using experimental electrical and thermal conductivity data both phonon ($\kappa_{ph}$) and electronic thermal conductivity ($\kappa_e$) were calculated. The values of $\kappa_{ph}$ for the (1-x)PbTe/(x)CoSb$_3$ composites are presented in Fig.7(b). Proportional changes in $\kappa_{ph}$ with an increasing volume fraction of CoSb$_3$ are observed. In Fig.7(c) $\kappa$ and $\kappa_{ph}$ in (0.50)PbTe/(0.50)CoSb$_3$ composite for a different particle sizes of CoSb$_3$ phase are almost identical. Additionally, in Fig.7(d), contributions from $\kappa_e$ and $\kappa_{ph}$ to total thermal conductivity for composites at 50°C are presented. For samples with x=0.00, 0.25, and 0.50, total thermal conductivity $\kappa$ is almost constant. However, due to the dominating nature of $\kappa_{ph}$ in the CoSb$_3$ phase, proportions of the $\kappa_{ph}$ rises with the increase of the amount of CoSb$_3$ phase.

A minute changes of $\kappa_{ph}$ in the (0.50)PbTe/(0.50)CoSb$_3$ composite with different particle sizes of CoSb$_3$ as well as the proportional increase in phonon thermal conductivity with an increase in the CoSb$_3$ phase fraction indicates that the phonon scattering at the boundary between PbTe and CoSb$_3$ phases is very small. To confirm this statement, we analyzed the obtained result in terms of acoustic impedance mismatch (AIM) model and interface thermal resistance ($R_{int}$).

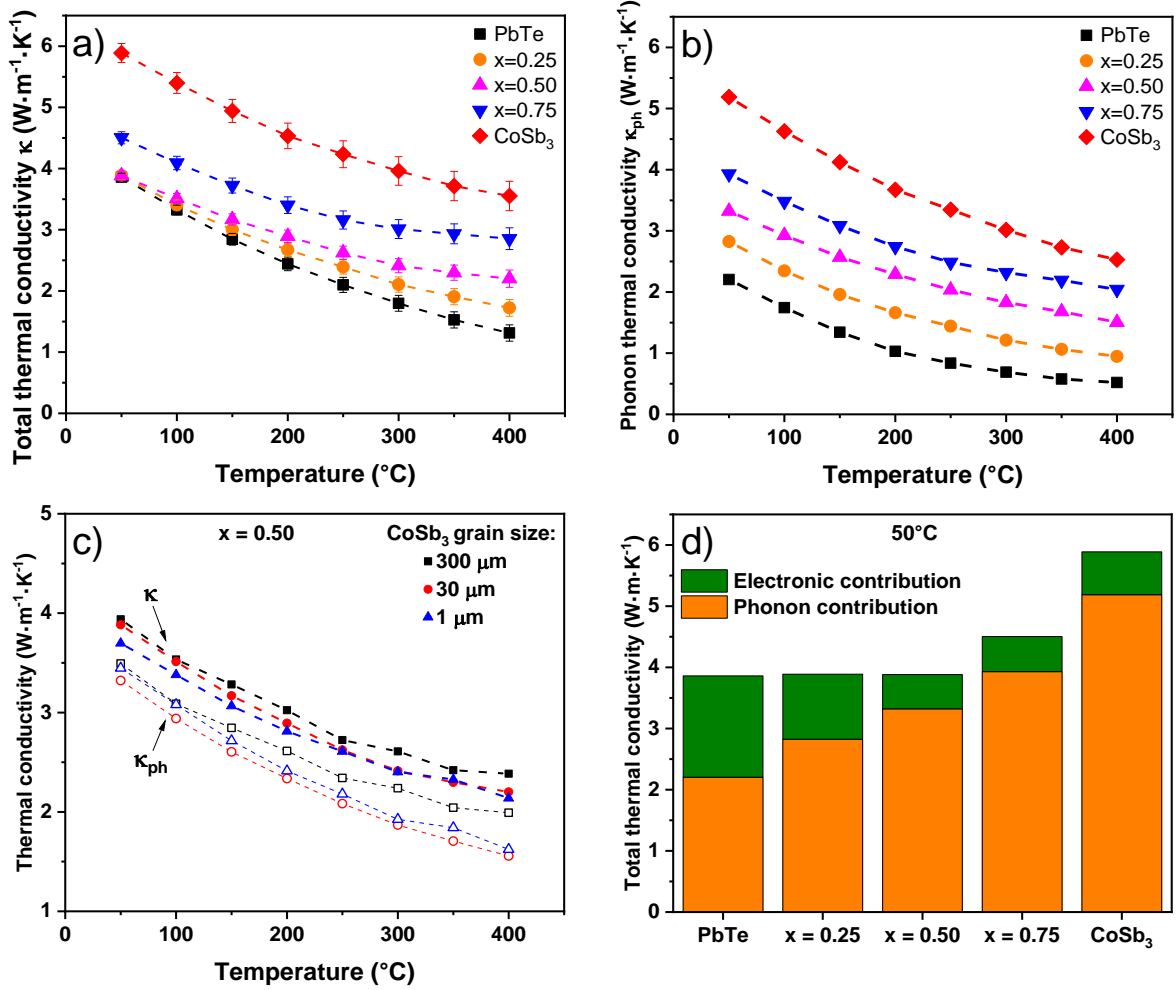

**Figure 7.** (a) Total thermal conductivity ($\kappa$), (b) phonon thermal conductivity ($\kappa_{ph}$) for (1-$x$)PbTe/($x$)CoSb$_3$, (0≤x≤1) composite with 30 µm particle size of CoSb$_3$. (c) $\kappa$ and $\kappa_{ph}$ for (0.50)PbTe/(0.50)CoSb$_3$ composite for a different particle size of CoSb$_3$ phase (dashed lines are a guide for an eye), (d) Phonon contribution ($\kappa_{ph}$) and electronic contribution ($\kappa_e$) to total thermal conductivity ($\kappa$) in (1-$x$)PbTe/($x$)CoSb$_3$ for 0≤x≤1.

### 4.2 ACOUSTIC IMPEDANCE MISMATCH (AIM) MODEL

When phonons inside the polycrystalline composite travel through one material and approaching the interface of the second material, some fraction of these phonons ($q$) will have the incident angle (with the surface of the junction) below a critical value, and they have a probability of getting transmitted through the barrier. The AIM model considers the phonons within the critical angle, and the fraction of phonons $q$ can be calculated using the following formula [27]:

$$q = \frac{1}{2}\left(\frac{v_m}{v_d}\right)^2 \quad (2)$$

where $v_m$ and $v_d$ are sound velocities of the matrix (PbTe) and dispersed phase (CoSb$_3$), respectively. AIM model assumes that the conduction of the heat is dominated by transverse acoustic phonons with frequencies corresponding to the linear regions of phonon dispersion curves. This assumption allows us to assign the measured sound velocities to the velocities of phonons [27]. Measured velocities are presented in Table II. We calculate the fraction of phonons ($q$) using the sound velocities measured only for transverse phonons and is found to be 0.18. For these phonons, we calculate the probability of their transmission and reflection at the interface [55,56]. First, we calculate the acoustic impedances ($Z$) for each of the joined materials:

$$Z_i = \rho_i \cdot v_i \qquad (3)$$

where $\rho_i$ is the density and $v_i$ is the sound wave velocity of the $i^{th}$ material. The probability of phonon reflection ($R_{1-2}$) and phonon transmission ($T_{1-2}$) between two materials is defined as:

$$R_{1-2} = \left(\frac{Z_1 - Z_2}{Z_1 + Z_2}\right)^2 \qquad (4)$$

$$T_{1-2} = \frac{4 Z_1 \cdot Z_2}{(Z_1 + Z_2)^2} \qquad (5)$$

According to this model, a large difference in the acoustic impedances will result in the enhanced probability of phonon reflection ($R_{1-2}$) at the interface and the reduced probability of phonon transmission ($T_{1-2}$). In the present study, acoustic impedance is calculated for both single-phase materials using measured values of density and sound velocity according to the Eq. (3) as well as both probabilities using Eq. (4) and (5) and are shown in Table II. Since the measurement of sound velocity on the flaw detector gives values for transverse and longitudinal directions separately, average values are presented for comparison of the obtained values. According to the AIM model, the average probability of phonon reflection on the boundary between PbTe and $CoSb_3$ phases is only 3.4%, which seems to agree with the proportional change of phonon thermal conductivity $\kappa_{ph}$ in the composite (Fig. 7). Further, we calculate the theoretical value of the interface thermal resistance ($R_{int}$) using the Debye model given by equation [27]:

$$R_{int}^D(T) = \frac{4}{\rho(T) \cdot c_p(T) \cdot v_D^{25} \cdot \eta^{25}} \qquad (6)$$

where $\rho(T)$ is the temperature dependence of matrix density obtained using thermal expansion coefficient [57], $c_p(T)$ is the temperature dependent specific heat of matrix obtained from the LFA measurement, $v_D^{25}$ is the Debye velocity [58] of the matrix at 25°C and $\eta^{25} = q \cdot T_{1-2} = 0.174$. The experimental values of $R_{int}$ on PbTe-$CoSb_3$ layered sample were determined using a double layer model incorporated in the software of the LFA apparatus [59].

**Table II.** Measured and calculated values of parameters necessary for the application of the AIM model.

| Material | $\rho$ [g/cm$^3$] | $v$ [m/s] | | $Z$ [kg/(m$^2$s)] | | $R_{AB}$ [%] (aver.) | $T_{AB}$ [%] (aver.) |
|---|---|---|---|---|---|---|---|
| | | Tr. | Lon. | Tr. | Lon. | | |
| PbTe | 7.98 | 1620 | 2960 | 13284 | 24272 | 3.4 | 96.6 |
| $CoSb_3$ | 7.26 | 2700 | 4510 | 20520 | 34276 | | |

The theoretically calculated values of $R_{int}^D$ and experimental results of $R_{int}$ are shown in Fig. 8. It is seen that both the measured and calculated values of $R_{int}$ are almost constant in the investigated temperature range. We estimated that the measured $R_{int}$ is only ~1% of the resistance of the whole sample, therefore we consider the compliance of both values within the one order of magnitude as satisfactory for further calculations.

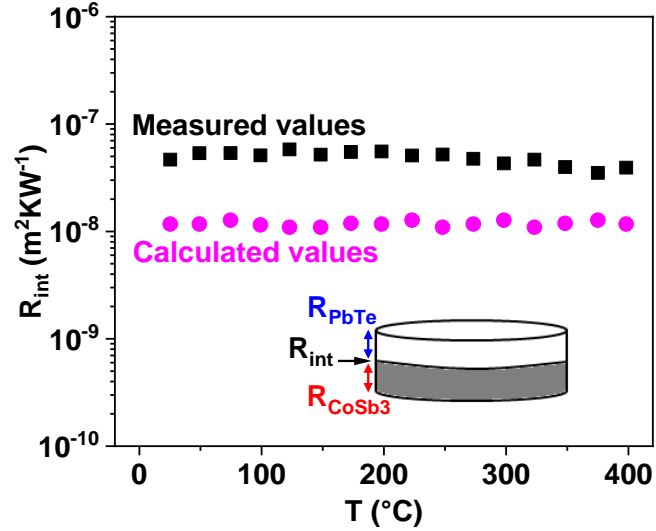

**Figure 8.** Experimental (squares) and theoretical (circles) interface thermal resistances in the PbTe/CoSb$_3$ layered sample.

### 4.3. EFFECTIVE MEDIA MODELS

For additional validation of the experimental results, we used effective media models for comparison. The Bruggeman asymmetrical model assumes one material to be a continuous matrix, and the second one to be in the form of spherical particles dispersed uniformly within this matrix. An essential feature of the Bruggeman asymmetrical model is the consideration of the presence of the $R_{int}$ between the matrix and the dispersed phase and is given by the formula [60]:

$$(1-x)^3 = \left(\frac{\kappa_m}{\kappa}\right)^{\frac{1+2\alpha}{1-\alpha}} \left(\frac{\kappa - \kappa_i(1-\alpha)}{\kappa_m - \kappa_i(1-\alpha)}\right)^{\frac{3}{1-\alpha}} \qquad (7)$$

where $\kappa_m$, $\kappa_i$, $\kappa$ are phonon thermal conductivity of the matrix, inclusion, and the whole composite, respectively. The inclusion volume phase fraction is given by $x$, and $\alpha = a/a_K$, where $a$ is the radius of the inclusions and $a_K (= R_{int} \cdot \kappa_m)$ is the Kapitza radius. Due to the mentioned asymmetry of this model, obtained values are valid up to a phase fraction of the dispersed phase equal to 0.50. Because we investigated the whole span of composite composition, we applied analogous calculation of theoretical values to both ends of the composition range. We consider this approach valid since both components play an interchangeable role in the conduction of heat, and both are in the form of similar particles. Dashed lines for $x$=0 - 0.50 in Fig. 9 represents the calculation results assuming PbTe as a matrix and CoSb$_3$ as a filler, and the dashed lines for $x$=0.50 – 1.00 represent an opposite case. We conclude that the $R_{int} = 6 \cdot 10^{-8}\ m^2KW^{-1}$ between the PbTe and CoSb$_3$ phases does not reduce the phonon thermal conductivity of the investigated composite material.

We also consider the rule of mixtures to depict the phonon thermal conductivity of the composite, shown by a straight dash-dotted line according to the equation:

$$\kappa = \kappa_1 \cdot x_1 + \kappa_2 \cdot (1 - x_1) \tag{8}$$

where $\kappa_1$, $\kappa_2$ are phonon thermal conductivities of phase 1 and 2 respectively, and $x_1$ is the volume phase fraction of phase 1. To verify which model fits the best to the experimental data in the (1-$x$)PbTe/($x$)CoSb$_3$ composite for 0≤$x$≤1, we performed statistical analysis for each model using the coefficient of determination presented in Table III. For each temperature, the highest $R^2$ is underlined to visualize which model fits best to the experimental data for given temperatures. For lower temperatures, the Bruggeman asymmetrical model fits better to the measured values; however, at higher temperatures, experimental data is better described by the rule of mixtures. It confirms the previous measurements and calculations of $R_{int}$ suggests that the obtained values of this parameter are reasonable. Moreover, it, suggests that the effect of interface thermal resistance in this composite is small when the particle size of the dispersed phase is ~30 µm.

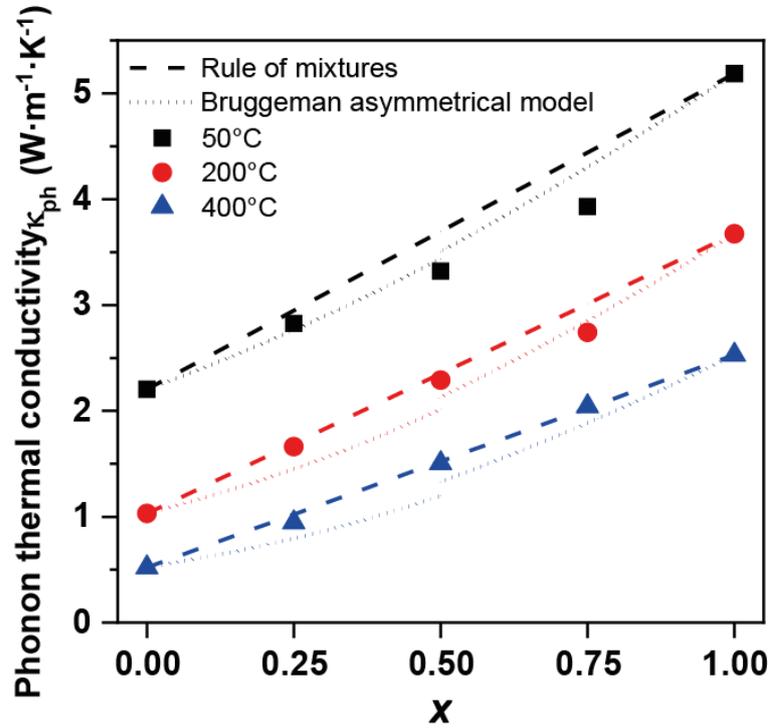

**Figure 9.** Comparison between experimental data (symbols) of phonon thermal conductivity ($\kappa_{ph}$) in the (1-$x$)PbTe/($x$)CoSb$_3$ for 0≤$x$≤1, composite with the Bruggeman asymmetrical model (dot lines), and the rule of the mixtures (dashed lines) for three different temperatures.

**Table III.** The coefficient of determination ($R^2$) obtained from the fitting of two models to the experimental data of composite thermal conductivity.

| Model | | 50°C | 200°C | 400°C |
|---|---|---|---|---|
| | | $R^2$ | | |
| Bruggeman asymmetrical | 0<x<0.50 | 0.972 | 0.847 | 0.756 |
| | 0.50<x<1.00 | 0.904 | 0.962 | 0.895 |
| | Average | <u>0.938</u> | 0.905 | 0.826 |
| Rule of mixtures | | 0.920 | <u>0.981</u> | <u>0.997</u> |

      The Bruggeman model allowed to predict the phonon thermal conductivity of the composite as a function of the particle size of the dispersed phase. Figure 10 shows the dependence of (1-$x$)PbTe/($x$)CoSb$_3$ composite phonon thermal conductivity vs. $x$ for different particle sizes of the CoSb$_3$ phase at two temperatures. It is seen that for the particle size of the dispersed, highly conductive CoSb$_3$ phase $a = 1.74 \cdot a_K \approx 230$ nm the phonon thermal conductivity $\kappa_{ph}$ of the composite is not dependent on its content. It is because the effective thermal conductivity of particles is equal to the conductivity of the PbTe matrix. The phonon thermal conductivity of the composite decreases, even below phonon thermal conductivity of the PbTe matrix, when the particle size of the dispersed CoSb$_3$ phase is smaller than $1.74 \cdot a_K$. As the particle size of the dispersed phase decreases, the relative contact area between phases increases, and the interface thermal resistance becomes more prominent [61,62]. The role of size effects on phonon thermal conductivity was also investigated for single-phase materials, e.g., diamond, silicon [63], or SrTiO$_3$ [64].

      Figure 11 presents theoretical values of (0.50)PbTe/(0.50)CoSb$_3$ composite phonon thermal conductivity vs. particle size of CoSb$_3$ at three different temperatures (dashed lines) obtained from the Bruggeman asymmetrical model along with the experimental data (squares). Our experimental results for different particle sizes are in agreement with the model, which predicts that for relatively big grains (1μm - 300μm), the phonon thermal conductivity of the composite remains almost constant. However, a significant decrease in composite phonon thermal conductivity is visible below 1μm. At Kapitza radius $a_K$, the thermal conductivity of composite is equal to the phonon thermal conductivity of the matrix (PbTe), and below $a_K$, composite phonon thermal conductivity is reduced even further. This indicates that utilization of correlation between $R_{int}$ and $a_K$ can lead to producing a thermoelectric composite with attractively lower thermal conductivity. This approach, along with improved or invariant power factor using band structure engineering of semiconductor-semiconductor heterojunctions, can be promising for preparing high performance thermoelectric composite materials.

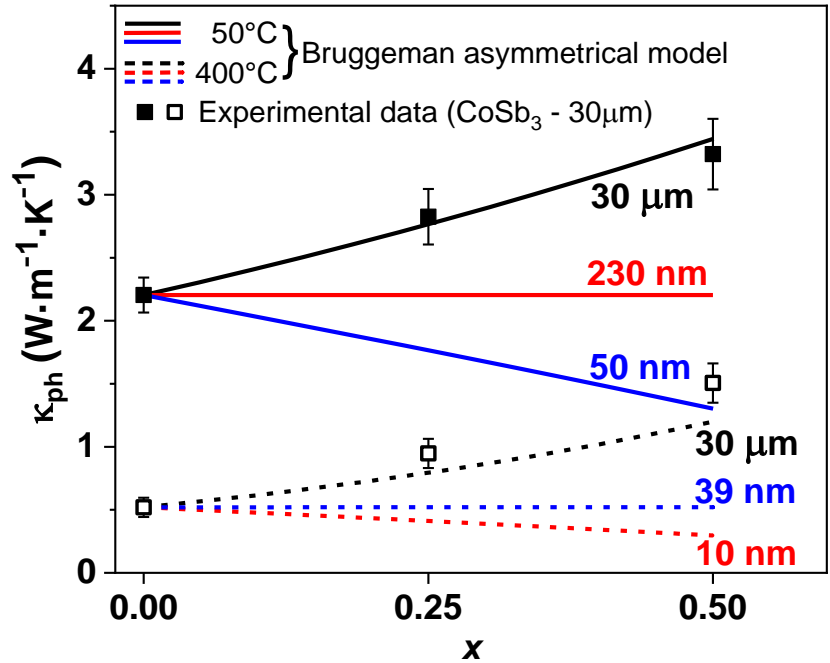

**Figure 10.** Phonon thermal conductivity ($\kappa_{ph}$) estimated using the Bruggeman asymmetrical model for different volume fractions for $CoSb_3$ phases (with different inclusion sizes) at two different temperatures. Experimental data obtained in this work are represented by symbols (full squares – 50°C, open squares – 400°C).

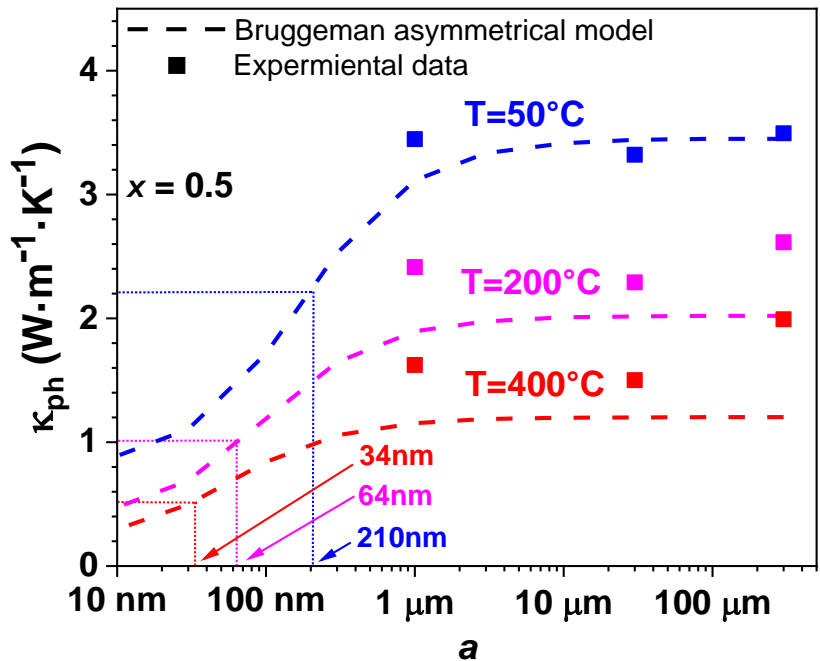

**Figure 11.** Theoretical $\kappa_{ph}$ of $(0.50)PbTe/(0.50)CoSb_3$ composite obtained from the Bruggeman asymmetrical model as a function of the $CoSb_3$ inclusion size (*a*). Horizontal lines represent the thermal conductivity of the PbTe phase (matrix in the Bruggeman model) at different temperatures, vertical lines represent the corresponding critical size of the $CoSb_3$ particles (inclusions).

5. CONCLUSIONS

Chemically and thermally stable composites made of *n*-type PbTe ($Pb_{0.99}Sb_{0.01}Te$) and n-type $CoSb_3$ ($CoSb_{2.94}Te_{0.06}$) thermoelectric materials are synthesized, and systematic investigation of their microstructural and thermal properties is performed. The X-ray diffraction analysis confirmed that only these phases co-exist in the composite. Chemical analysis (EDXS) does not show any significant sign of chemical diffusion of elements between the two phases and supports conclusions from the XRD analysis. A monotonous increase in the phonon thermal conductivity $\kappa_{ph}$ of the composite with the increase of the $CoSb_3$ phase fraction $x$ is observed. This behavior of $\kappa_{ph}$ is well explained by the effective media theory using the interface thermal resistance $R_{int}$ and mismatch in acoustic impedance ($Z_{PbTe}$ and $Z_{CoSb_3}$) between both the phases. The phonon thermal conductivity $\kappa_{ph}$ of the composite, as a function of the $CoSb_3$ phase fraction $x$, follows the Bruggeman asymmetrical model better than the simple rule of mixtures at higher temperatures. We show that the phonon thermal conductivity of the composite can be reduced below the phonon thermal conductivity of the PbTe matrix when the particle size of the dispersed $CoSb_3$ phase is smaller than the $1.74 \cdot a_K$ (~230 nm at 320K). This critical inclusion size is found to be decreasing with temperature (~40 nm at 700K). It is attributed to the reduction in the phonon thermal conductivity of the PbTe matrix at higher temperatures due to the decreasing mean free path of phonons.

This study confirms that effective media theory is useful for the prediction of thermal transport in composites made of two thermoelectric materials. The results obtained in the present research shows new paths in the advanced engineering of thermoelectric composite materials by tuning acoustic impedance mismatch (i.e differences in elastic properties of components) and using optimized particle size of the dispersed phase.

6. CONFLICT OF INTEREST

There are no conflicts to declare.

7. ACKNOWLEDGMENTS

We are very grateful to Prof. Yuri Grin for valuable comments regarding the research and support. This research was financed by the 'New approach for development of efficient materials for direct conversion of heat into electricity' TEAM-TECH/2016-2/14 project which is carried out within the TEAM-TECH program of the Foundation for Polish Science co-financed by the European Union under the European Regional Development Fund and the beneficiary of this project is The Lukasiewicz Research Network - The Cracow Institute of Technology (Poland).

Systems Utilising Thermoelectric Generators and Heat Pipes. *Applied Thermal Engineering*. Elsevier Ltd May 25, 2016, pp 490–495. https://doi.org/10.1016/j.applthermaleng.2015.10.081.